\documentclass[a4paper,12pt]{article}
\usepackage{amssymb}

\input{tcilatex}

\begin{document}

\setcounter{page}{0} \topmargin0pt \oddsidemargin5mm \renewcommand{%
\thefootnote}{\fnsymbol{footnote}} \newpage \setcounter{page}{0} 
\begin{titlepage}
\begin{flushright}
Berlin Sfb288 Preprint  \\
hep-th/0112097
\end{flushright}
\vspace{0.2cm}
\begin{center}
{\Large {\bf Mutually local fields from form factors} }

\vspace{0.8cm}
{\large  O.A.~Castro-Alvaredo  and  A.~Fring }

\vspace{0.2cm}
{Institut f\"ur Theoretische Physik, 
Freie Universit\"at Berlin,\\
Arnimallee 14, D-14195 Berlin, Germany }
\end{center}
\vspace{0.5cm}
 
\renewcommand{\thefootnote}{\arabic{footnote}}
\setcounter{footnote}{0}

\begin{abstract}

We compare two different methods of computing form factors. One is the well established
procedure of solving the form factor consistency equations and the other is to
represent the field content as well as the particle creation operators
in terms of fermionic Fock operators. We compute the corresponding matrix elements 
for the complex free fermion and the Federbush model. 
The matrix elements only satisfy the form factor 
consistency equations involving anyonic factors of local commutativity when the corresponding
operators are local. We carry out the ultraviolet limit, analyze the
momentum space cluster properties and demonstrate how the Federbush model can
be obtained from the  $SU(3)_3$-homogeneous sine-Gordon model. We propose a new 
class of Lagrangians which  constitute a generalization of the Federbush model in
a Lie algebraic fashion.  For these models we evaluate the associated scattering matrices 
from first principles, 
which can alternatively also be obtained in a certain limit of the homogeneous sine-Gordon models.
\end{abstract}
\vfill{ \hspace*{-9mm}
\begin{tabular}{l}
\rule{6 cm}{0.05 mm}\\
Based on talks presented at the conferences:\\
``From QCD to integrable models, old results and new developments"\\
(Nor Amberd, Armenia, September, 2001);\\

``APCTP - Nankai joint symposium on lattice statistics and mathematical physics"\\
(Tianjin, China, October, 2001);\\

``ICMS workshop on classical and quantum integrable systems and their symmetries"\\
(Edinburgh, Scottland, December, 2001)

\end{tabular}}
\end{titlepage}
\newpage

\section{Introduction}

One of the most central concepts in relativistic quantum field theory, like
Einstein causality and Poincar\'{e} covariance, are captured in local field
equations and commutation relations. In fact this principle is widely
considered as so pivotal that it constitutes the base of a whole subject,
i.e. local quantum physics (algebraic quantum field theory) \cite{Haag}
which takes the collection of all operators localized in a particular region
generating a von Neumann algebra, as its very starting point.

On the other hand, in the formulation of a quantum field theory, one may
alternatively start from a particle picture and investigate the
corresponding scattering theories. In particular for 1+1 dimensional
integrable quantum field theories this latter approach has been proved to be
impressively successful. As its most powerful tool one exploits here first
the bootstrap principle \cite{boot1,boot2,boot3}, which allows to write down
exact, i.e. non-perturbative, scattering matrices. Ignoring subtleties of
non-asymptotic states, it is essentially possible to obtain the particle
picture from the field formulation by means of the LSZ-reduction formalism 
\cite{LSZ}. However, the question of how to reconstruct the field content,
or at least part of it, from the scattering theory is in general still an
outstanding issue.

In the context of 1+1 dimensional integrable quantum field theories the
identification of the operators is based on the assumption, dating back to
the initial papers \cite{Kar}, that each solution to the form factor
consistency equations \cite{Kar,Smir,YZam,BFKZ} corresponds to a particular
local operator. Consequently one approach, as outlined in section 3.1., to
construct the quantum field theory consists of solving systematically this
set of equations and thereafter pin down the nature of the operator. To do
this, numerous authors have used diverse arguments. For instance, most
conventional, one may study the asymptotic behaviours or perform
perturbation theory. More in the spirit of an exact formulation is to take
symmetries into account and to formulate quantum equations of motion or
conservation laws \cite
{YZam,Zamocorr,BFKZ,deter,KM,Smir2,CardyM,BK,CFK,CF1,CF2,CF3,CF5}. However,
these computations can not really be regarded as a stringent identification,
since they only relate particular solutions to each other and lack
systematics. Even when taking them as a mere consistency check one should be
cautious, since such equations also hold for matrix elements which do not
satisfy the consistency equations, as argued in section 5.2 in more detail.
An approach with somewhat more underlying systematic is to carry out the
ultraviolet limit of the theory and appeal to the well understood
classification scheme of conformal field theory \cite
{CardyM,Smir2,DSC,CFK,CF1,CF2,CF3}. Naming the operators in the massive
model is then in one-to-one correspondence with the conformal field theory.
So far it is still problematic here to unravel degeneracies \cite{CF1}.
Furthermore, one should be cautious when using this correspondence, since
there might be operators, so-called ``shadow operators'', in the massive
model which do not possess a counterpart in the underlying conformal field
theory \cite{shadow}.

This talk is also devoted to this question in the sense that we provide
explicit expressions for operators $\mathcal{O}(x)$ located at $x$ in terms
of fermionic Fock fields. Particular emphasis is put on the question whether
these operators are really local in the sense that they (anti)-commute for
space-like separations with themselves, 
\begin{equation}
\lbrack \mathcal{O}(x),\mathcal{O}(y)]=0\qquad \text{for}\qquad (x-y)^{2}<0
\label{loc}
\end{equation}
and how this property is reflected in the form factor consistency equations.
It will turn out that from possible matrix elements the form factor
consistency equations select out those which correspond to mutually local
operators. We argue that the presence of the factor of local commutativity
in these equations is absolutely essential.

\section{Prerequisites}

The fundamental observation, on which all further analysis hinges, is that
integrability, which means here the existence, one does not need to know its
explicit form, of at least one non-trivial conserved charge, in 1+1
space-time dimensions implies the factorization of the n-particle scattering
matrix into a product of two-particle scattering matrices 
\begin{equation}
Z_{\mu _{n}}^{\dagger }(\theta _{n})\ldots Z_{\mu _{1}}^{\dagger }(\theta
_{1})\left| 0\right\rangle _{\text{out}}=\!\!\!\prod\limits_{1\leq i<j\leq
n}\!\!\!S_{\mu _{i}\mu _{j}}(\theta _{ij})Z_{\mu _{1}}^{\dagger }(\theta
_{1})\ldots Z_{\mu _{n}}^{\dagger }(\theta _{n})\left| 0\right\rangle _{%
\text{in}}\,.  \label{fact}
\end{equation}
As common we parameterize the two-momentum $\vec{p}$ by the rapidity
variable $\theta $ as $\vec{p}=m(\cosh \theta ,\sinh \theta )$ and
abbreviate $\theta _{ij}:=\theta _{i}-\theta _{j}$. The $Z_{\mu }^{\dagger
}(\theta )$ denote creation operators for stable particles of type $\mu $
with rapidity $\theta $, which obey the Faddev-Zamolodchikov algebra \cite
{FZ1,FZ2} 
\begin{equation}
Z_{i}^{\dagger }(\theta _{i})Z_{j}^{\dagger }(\theta _{j})=S_{ij}(\theta
_{ij})Z_{j}^{\dagger }(\theta _{j})Z_{i}^{\dagger }(\theta _{i})=\exp [2\pi
i\delta _{ij}(\theta _{ij})]Z_{j}^{\dagger }(\theta _{j})Z_{i}^{\dagger
}(\theta _{i})\,.  \label{ZF}
\end{equation}
As indicated in equation (\ref{ZF}), the two-particle scattering matrix $%
S_{ij}(\theta _{ij})$ can be expressed as a phase.

The basic assumption of the bootstrap program is now that every solution to
the unitary-analyticity, crossing and fusing bootstrap equations\footnote{%
For the purpose of this talk we suppose that there is no backscattering in
the theory such that the Yang-Baxter equation constitutes no constraint.} 
\begin{equation}
S_{ij}(\theta )=S_{ji}(-\theta )^{-1}=S_{j\bar{\imath}}(i\pi -\theta ),\quad
\prod_{l=i,j,k}S_{dl}(\theta +i\eta _{l})=1\,\,\,,  \label{boot}
\end{equation}
($\eta _{l}\in \Bbb{Q}$ are the fusing angles encoding the mass spectrum and
the anti-particle of $i$ is $\bar{\imath}$), which admits a consistent
explanation of all poles inside the physical sheet (that is $0<\func{Im}%
\theta <\pi $), leads to a local quantum field theory. There exists no
rigorous proof for this assumption, however, it is supported by numerous
explicitly constructed examples.

In order to pass from scattering theory to fields, we want to determine the
form factors, i.e. the matrix element of a local operator $\mathcal{O}(x)$
located at the origin between a multi-particle in-state and the vacuum 
\begin{equation}
F_{n}^{\mathcal{O}|\mu _{1}\ldots \mu _{n}}(\theta _{1},\ldots ,\theta
_{n})\equiv \left\langle \mathcal{O}(0)\mathcal{\,}Z_{\mu _{1}}^{\dagger
}(\theta _{1}), \ldots, Z_{\mu _{n}}^{\dagger }(\theta _{n})\right\rangle _{%
\text{in}}\,.  \label{1}
\end{equation}
We distinguish here between the mere matrix element $\tilde{F}_{n}^{\mathcal{%
O}}$ and the particular ones which also solve the consistency equations as
stated in section 3.1. In that case we denote them as $F_{n}^{\mathcal{O}}$.

\section{Determination of form factors}

\subsection{Solving the consistency equations}

Various schemes have been suggested to compute the objects in equation (\ref
{1}). One of the original approaches is modeled in spirit closely on the set
up for the determination of exact scattering matrices. It consists of
solving a system of consistency equations which have to hold for the
n-particle form factors based on some natural physical assumptions, like
unitarity, crossing and bootstrap fusing properties \cite{Kar,Smir,YZam,BFKZ}
\begin{eqnarray}
F_{n}^{\mathcal{O}|\ldots \mu _{i}\mu _{j}\ldots }(\ldots ,\theta
_{i},\theta _{j},\ldots ) &=&F_{n}^{\mathcal{O}|\ldots \mu _{j}\mu
_{i},\ldots }(\ldots ,\theta _{j},\theta _{i},\ldots )S_{\mu _{i}\mu
_{j}}(\theta _{ij})\,,  \label{W1} \\
F_{n}^{\mathcal{O}|\mu _{1}\ldots \mu _{n}}(\theta _{1}+2\pi i,\ldots
,\theta _{n}) &=&\gamma _{\mu _{1}}^{\mathcal{O}}\,F_{n}^{\mathcal{O}|\mu
_{2}\ldots \mu _{n}\mu _{1}}(\theta _{2},\ldots ,\theta _{n},\theta
_{1})\,\,,  \label{W2} \\
F_{n}^{\mathcal{O}|\mu _{1}\ldots \mu _{n}}(\theta _{1}+\lambda ,\ldots
,\theta _{n}+\lambda ) &=&e^{s\lambda }F_{n}^{\mathcal{O}|\mu _{1}\ldots \mu
_{n}}(\theta _{1},\ldots ,\theta _{n})\,\,,  \label{rel} \\
\limfunc{Res}_{{\small \bar{\theta}}\rightarrow {\small \theta }_{0}}{\small %
F}_{n+2}^{\mathcal{O}|\bar{\mu}\mu \mu _{1},\ldots ,\mu _{n}}{\small (\bar{%
\theta}+i\pi ,\theta }_{0}\!\!\!\! &&\!\!\!\!{\small ,\theta }_{1}{\small %
\ldots \theta }_{n}{\small )}=i(1-\gamma _{\mu }^{\mathcal{O}%
}\prod_{l=1}^{n}S_{\mu \mu _{l}}(\theta _{0l}))  \nonumber \\
&&\qquad \quad \qquad \quad \times {\small F}_{n}^{\mathcal{O}|\mu
_{1}\ldots \mu _{n}}{\small (\theta }_{1}{\small ,\ldots ,\theta }_{n}%
{\small ).}  \label{kin}
\end{eqnarray}
Here $s$ is the Lorentz spin of the operator $\mathcal{O}$ and $\lambda $ is
an arbitrary real number. We omitted here the so-called bound state residue
equation, which relates an $(n+1)$- to an $n$-particle form factor, since it
will be of no importance to the explicit models we consider here. We stress
the importance of the constant $\gamma _{\mu }^{\mathcal{O}}$, the factor of
so-called local commutativity defined through the equal time exchange
relation of the local operator $\mathcal{O}(x)$ and the field $\mathcal{O}%
_{\mu }(y)$ associated to the particle creation operators $Z_{\mu }^{\dagger
}(\theta )$%
\begin{equation}
\mathcal{O}_{\mu }(x)\mathcal{O}(y)=\gamma _{\mu }^{\mathcal{O}}\,\mathcal{O}%
(y)\,\mathcal{O}_{\mu }(x)\qquad \,\text{for\thinspace \thinspace\ \ \ }%
x^{1}>y^{1}\,,  \label{local}
\end{equation}
with $x^{\mu }=(x^{0},x^{1})$. This factor $\gamma _{\mu }^{\mathcal{O}}$,
which appeared for the first time in \cite{YZam} is very often omitted in
the analysis or simply taken to be one, but it can be seen that already in
the Ising model it is needed to set up the equations consistently \cite{YZam}%
. We also emphasize that this factor is not identical to the statistics
factor, associated to an exchange of particles, which is sometimes extracted
explicitly from the scattering matrix, see e.g. \cite{BFKZ}. This factor
carries properties of the operator and not just of the $Z^{\dagger }$'s. An
immediate consequence of its presence is that a frequently made statement
has to be revised, namely, that (\ref{W1})-(\ref{kin}) constitute operator
independent equations, which require as the only input the two-particle
scattering matrix. Here we demonstrate that apart from $\pm 1$, which
already occur in the literature, this factor can be a non-trivial phase.
Thus the form factor consistency equations contain also explicitly
non-trivial properties of the operators.

To solve these equations at least for the lowest $n$-particle form factors
is a fairly well established procedure, but it still remains a challenge to
find closed analytic solutions for all $n$-particle form factors. We briefly
recall the principle steps of the general solution procedure. For any local
operator $\mathcal{O}$ one may anticipate the pole structure of the form
factors and extract it explicitly in form of factorizing an ansatz. This
might turn out to be a relatively involved matter due to the occurrence of
higher order poles in some integrable theories, but nonetheless it is always
possible. Thereafter the task of finding solutions may be reduced to the
evaluation of the so-called minimal form factors and to solving a (or two if
bound states may be formed in the model) recursive equation for a polynomial
which results from (\ref{kin}) with the mentioned ansatz. The first task can
be carried out relatively easily, especially if the related scattering
matrix is given as a particular integral representation \cite{Kar}. The
second task is rather more complicated and the heart of the whole problem in
this approach. Having a seed for the recursive equation, that is the lowest
non-vanishing form factor one can in general compute from them several form
factors which involve more particles. (This seed could be either a known
form factor when the model reduces to some solved case or possibly the
vacuum expectation value of the operator, which is not known in most cases.)
Unfortunately, the equations become relatively involved after several steps.
Aiming at the solution for all $n$-particle form factors, it is therefore
highly desirable to unravel a more generic structure which enables one to
formulate rigorous proofs. Several examples \cite
{Zamocorr,deter,CFK,CF1,CF2,CF3} have shown that often the general solution
may be cast into the form of determinants whose entries are elementary
symmetric polynomials. Presuming such a structure which, at present, may be
obtained by extrapolating from lower particle solutions to higher ones or by
some inspired guess, one can rigorously formulate proofs such solutions.
These determinant expressions allow directly to write down equivalent
integral representations, see e.g. \cite{CFK}. There exist also different
types of universal expressions like for instance the integral
representations presented in \cite{Smir,BFKZ}. However, these type of
expressions are sometimes only of a very formal nature since to evaluate
them concretely for higher $n$-particle form factors requires still a
considerable amount of computational effort.

\subsection{Direct computation of matrix elements}

The most direct way to compute the matrix elements in (\ref{1}) is to find
explicit representations for the operators $Z_{\mu }^{\dagger }(\theta )$
and $\mathcal{O}(x)$. For instance in the context of lattice models this is
a rather familiar situation and one knows how to compute matrix elements of
the type (\ref{1}) directly. The problem is then reduced to a purely
computational task (albeit non-trivial), which may, for instance, be solved
by well-known techniques of algebraic Bethe ansatz type, see e.g. \cite
{Bethe}. In the context of field theory a similar way of attack to the
problem has been followed by exploiting a free field representation for the
operators $Z_{\mu }^{\dagger }(\theta )$ and $\mathcal{O}(x)$, in form of
Heisenberg algebras or their q-deformed version. So far a successful
computation of the n-particle form factors with this approach is limited to
a rather restricted set of models and in particular for the sine-Gordon
model, which is a model extensively studied by means of other approaches 
\cite{Smir,BFKZ,BK}, only the free Fermion point can be treated successfully
at present \cite{MJ,Pak,LZ}. One of the purposes of this talk is to advocate
another approach, namely the evaluation of the matrix elements (\ref{1})
based on an expansion of the operators in the conventional fermionic Fock
space. Recalling the well-known fact that in 1+1 space-time dimensions the
notions of spin and statistics are not intrinsic, it is clear that both
approaches are equally legitimate.

So, how do we represent the operators $Z_{\mu }^{\dagger }(\theta )$ and $%
\mathcal{O}(x)$? For the former this task is solved. A representation for
these operators in the bosonic Fock space was first provided in \cite{AF} 
\begin{equation}
Z_{i}^{\dagger }(\theta )=\exp \left[ -i\int_{\theta }^{\infty }d\theta
^{\prime }\,\delta _{il}(\theta -\theta ^{\prime })a_{l}^{\dagger }(\theta
^{\prime })a_{l}(\theta ^{\prime })\right] a_{i}^{\dagger }(\theta )\,.
\label{real}
\end{equation}
By replacing a constant phase with the rapidity dependent phase $\delta
_{ij}(\theta )$ and turning the expression into a convolution with an
additional sum over $l$, the expression (\ref{real}) constitutes a
generalization of formulae found in the late seventies \cite{spin}, which
interpolate between bosonic and fermionic Fock spaces for arbitrary spin.
The latter construction may be viewed as a continuous version of a
Jordan-Wigner transformation \cite{JW}, albeit on the lattice the
commutation relations are not purely bosonic or fermionic, since certain
operators anti-commute at the same site but commute on different sites.
Alternatively, one may also replace the bosonic $a$'s in (\ref{real}) by
operators satisfying the usual fermionic anti-commutation relations 
\begin{equation}
\left\{ a_{i}(\theta ),a_{j}(\theta ^{\prime })\right\} =0\qquad \text{%
and\qquad }\left\{ a_{i}(\theta ),a_{j}^{\dagger }(\theta ^{\prime
})\right\} =2\pi \delta _{ij}\delta (\theta -\theta ^{\prime })\,
\label{ferm}
\end{equation}
and note that the exchange relations (\ref{ZF}) are still satisfied \cite
{Notas}. In the following we want to work with this fermionic representation
of the FZ-algebra. Having obtained a fairly simple realization for the $Z$%
-operators, we may now seek to represent the operator content of the theory
in the same space. How to do this is not known in general and we have to
resort to a study of explicit models at this stage.

\section{Complex free Fermions}

Let us consider $N$ complex (Dirac) free Fermions described as usual by the
Lagrangian density 
\begin{equation}
\mathcal{L}_{\text{FF}}=\sum\nolimits_{\alpha =1}^{N}\bar{\psi}_{\alpha
}(i\gamma ^{\mu }\partial _{\mu }-m_{\alpha })\psi _{\alpha }\,.  \label{fff}
\end{equation}
We define a prototype auxiliary field 
\begin{eqnarray}
\chi _{\kappa }^{\alpha }(x) &=&\int \frac{d\theta d\theta ^{\prime }}{4\pi
^{2}}\left[ \kappa ^{\alpha }(\theta ,\theta ^{\prime })\left( a_{\alpha
}^{\dagger }(\theta )a_{\bar{\alpha}}^{\dagger }(\theta ^{\prime
})e^{i(p+p^{\prime })\cdot x}+a_{\alpha }(\theta )a_{\bar{\alpha}}(\theta
^{\prime })e^{-i(p+p^{\prime })\cdot x}\right) \right.  \nonumber \\
&&\!\!\!\!\!\!\!\left. +\kappa ^{\alpha }(\theta ,\theta ^{\prime }-i\pi
)\left( a_{\bar{\alpha}}^{\dagger }(\theta )a_{\bar{\alpha}}(\theta ^{\prime
})e^{i(p-p^{\prime })\cdot x}-a_{\alpha }(\theta )a_{\alpha }^{\dagger
}(\theta ^{\prime })e^{-i(p-p^{\prime })\cdot x}\right) \right]  \label{aux}
\end{eqnarray}
and intend to compute the matrix element of general operators composed out
of these fields 
\begin{equation}
\mathcal{O}^{\chi _{\kappa }^{\alpha }}(x)=\mathbf{:}e^{\chi _{\kappa
}^{\alpha }(x)}\mathbf{:},\;\hat{\mathcal{O}}^{\chi _{\kappa }^{\alpha }}(x)=%
\mathbf{:}\int \frac{dp_{\alpha }^{1}}{2\pi p_{\alpha }^{0}}(a_{\alpha
}(p)e^{-ip_{\alpha }\cdot x}+a_{\bar{\alpha}}^{\dagger }(p)e^{ip_{\alpha
}\cdot x})e^{\chi _{\kappa }^{\alpha }(x)}\mathbf{:}.  \label{aux1}
\end{equation}
Employing Wick's first theorem, we compute \cite{CF5} 
\begin{eqnarray}
\tilde{F}_{2n}^{\mathcal{O}^{\chi _{\kappa }^{\alpha }}|n\times \bar{\alpha}%
\alpha }(\theta _{1},\ldots ,\theta _{2n}) &=&\int \frac{d\theta
_{1}^{^{\prime }}\ldots d\theta _{2n}^{^{\prime }}}{n!}\prod_{i=1}^{n}\kappa
^{\alpha }(\theta _{2i-1}^{\prime },\theta _{2i}^{\prime })\det \mathcal{D}%
^{2n}\,\,,  \label{even} \\
\tilde{F}_{2n+1}^{\hat{\mathcal{O}}^{\chi _{\kappa }^{\alpha }}|\alpha
,n\times \bar{\alpha}\alpha }(\theta _{1},\ldots ,\theta _{2n+1}) &=&\int 
\frac{d\theta _{1}^{^{\prime }}\ldots d\theta _{2n+1}^{^{\prime }}}{n!}%
\prod_{i=1}^{n}\kappa ^{\alpha }(\theta _{2i}^{\prime },\theta
_{2i+1}^{\prime })\det \mathcal{D}^{2n+1},\,\,\;  \label{odd}
\end{eqnarray}
where $\mathcal{D}^{\ell }$ is a rank $\ell $ matrix whose entries are given
by 
\begin{equation}
\mathcal{D}_{ij}^{\ell }=\cos ^{2}[(i-j)\pi /2]\delta (\theta _{i}^{^{\prime
}}-\theta _{j})\,,\qquad 1\leq i,j\leq \ell \,.
\end{equation}
Note that $\mathcal{O}^{\chi _{\kappa }^{\alpha }}(x)$ and $\hat{\mathcal{O}}%
^{\chi _{\kappa }^{\alpha }}(x)$ are in general non-local operators in the
sense of (\ref{loc}). At the same time $\tilde{F}_{n}^{\mathcal{O}}$ is just
the matrix element as defined on the r.h.s. of (\ref{1}) and not yet a form
factor of a local field, in the sense that it satisfies the consistency
equations (\ref{W1})-(\ref{kin}), which imply locality of $\mathcal{O}$. A
rigorous proof of this latter implication to hold in generality is still an
open issue. Let us now specify the function $\kappa $. The free fermionic
theory possesses some very distinct fields, namely the disorder and order
fields 
\begin{equation}
\mu _{\alpha }(x)=\mathbf{:}e^{\omega _{\alpha }(x)}\mathbf{:\qquad }\text{%
and \qquad }\sigma _{\alpha }(x)=\mathbf{:}\hat{\psi}_{\alpha }(x)\mu
_{\alpha }(x)\mathbf{:},\quad \alpha =1,2,  \label{ggg}
\end{equation}
respectively. We introduced here the fields 
\begin{equation}
\omega _{\alpha }(x)=\chi _{\kappa }^{\alpha }(x),\quad \kappa ^{1}(\theta
,\theta ^{\prime })=-\kappa ^{2}(-\theta ,-\theta ^{\prime })=\frac{i}{2}%
\frac{e^{-\frac{1}{2}(\theta -\theta ^{\prime })}}{\cosh \frac{1}{2}(\theta
-\theta ^{\prime })}\quad \,.  \label{k}
\end{equation}
We compute \cite{CF5} the integrals in (\ref{even}) and (\ref{odd}) for this
case and obtained a closed expression for the n-particle form factors of the
disorder and order operators 
\begin{eqnarray}
F_{2n}^{\mu _{1}|n\times \bar{1}1}(\theta _{1},\ldots ,\theta _{2n})
&=&(-1)^{n}F_{2n}^{\mu _{2}|n\times \bar{2}2}(-\theta _{1},\ldots ,-\theta
_{2n})  \nonumber \\
F_{2n}^{\mu _{\bar{1}}|n\times \bar{1}1}(-\theta _{1},\ldots ,-\theta _{2n})
&=&(-1)^{n}F_{2n}^{\mu _{\bar{2}}|n\times \bar{2}2}(\theta _{1},\ldots
,\theta _{2n})  \nonumber \\
&=&i^{n}2^{n-1}\sigma _{n}(\bar{x}_{1},\bar{x}_{3},\ldots ,\bar{x}_{2n-1})%
\mathcal{B}_{n,n}\,,  \label{555} \\
F_{2n+1}^{\sigma _{1}|1(n\times \bar{1}1)}(\theta _{1},\ldots ,\theta
_{2n+1}) &=&(-1)^{n}F_{2n+1}^{\sigma _{2}|2(n\times \bar{2}2)}(-\theta
_{1},\ldots ,-\theta _{2n+1})  \nonumber \\
F_{2n+1}^{\sigma _{\bar{1}}|1(n\times \bar{1}1)}(-\theta _{1},\ldots
,-\theta _{2n+1}) &=&(-1)^{n}F_{2n+1}^{\sigma _{\bar{2}}|2(n\times \bar{2}%
2)}(\theta _{1},\ldots ,\theta _{2n+1})  \nonumber \\
&=&i^{n}2^{n-1}\sigma _{n}(\bar{x}_{1},\ldots ,\bar{x}_{2n-1})\mathcal{B}%
_{n,n+1},  \label{6}
\end{eqnarray}
with 
\begin{equation}
\mathcal{B}_{n,m}=\frac{\prod\nolimits_{1\leq i<j\leq n}(\bar{x}_{2i-1}^{2}-%
\bar{x}_{2j-1}^{2})\prod\nolimits_{1\leq i<j\leq m}(x_{2i}^{2}-x_{2j}^{2})}{%
\prod\nolimits_{1\leq i<j\leq n+m}(u_{i}+u_{j})}\,.  \label{habibi}
\end{equation}
Associated with the particles and anti-particles we introduced here the
quantities $x_{i}=\exp (\theta _{i})$ and $\bar{x}_{i}=\exp (\bar{\theta}%
_{i})$, respectively. The variable $u_{i}$ can be either of them. We also
employed the elementary symmetric polynomials $\sigma _{k}(x_{1},\ldots
,x_{n})$. The remaining form factors are zero due to the U(1)-symmetry of
the Lagrangian. One may easily verify that the expressions (\ref{555}) and (%
\ref{6}) indeed satisfy the consistency equations (\ref{W1})-(\ref{kin})
with $\gamma _{\bar{\alpha}}^{\mu _{\alpha }}=-1$ and $\gamma _{\bar{\alpha}%
}^{\sigma _{\alpha }}=1$ for $\alpha =1,2$. We also compute \cite{CF5} the
form factors associated to the trace of the energy-momentum tensor 
\begin{equation}
F_{2}^{T_{\;\;\mu }^{\mu }|\bar{\alpha}\alpha }(\theta ,\tilde{\theta}%
)=F_{2}^{T_{\;\;\mu }^{\mu }|\alpha \bar{\alpha}}(\theta ,\tilde{\theta}%
)=-2\pi im_{\alpha }^{2}\sinh \frac{\theta -\tilde{\theta}}{2}\,,
\end{equation}
which plays a distinct role in the ultraviolet limit.

We want to conclude this section with a general comment on the comparison
between the generic operators of the type (\ref{aux}), (\ref{aux1}) with
some general expressions for ``local'' operators which appear in the
literature \cite{Lask,Notas,Bert}. We carry out this argument in generality
without restriction to a concrete model. Let us restore in equation (\ref{1}%
) the space-time dependence, multiply the equation from the left with the
bra-vector $\left\langle Z_{\mu _{n}}^{\dagger }(\theta _{n})\ldots Z_{\mu
_{1}}^{\dagger }(\theta _{1})\right| $ and introduce the necessary amount of
sums and integrals over the complete states such that one can identify the
identity operator $\Bbb{I}$%
\begin{eqnarray*}
&&\!\!\sum\Sb n=1\ldots \infty  \\ \mu _{1}\ldots \mu _{n}  \endSb %
\int\limits_{-\infty }^{\infty }\frac{d\theta _{1}\ldots d\theta _{n}}{%
n!(2\pi )^{n}}F_{n}^{\mathcal{O}|\mu _{1}\ldots \mu _{n}}(\theta _{1}\ldots
\theta _{n})\left\langle Z_{\mu _{n}}^{\dagger }(\theta _{n})\ldots Z_{\mu
_{1}}^{\dagger }(\theta _{1})\right| e^{-i\sum_{j}p_{j}\cdot x} \\
&=&\!\!\sum\Sb n=1\ldots \infty  \\ \mu _{1}\ldots \mu _{n}  \endSb %
\int\limits_{-\infty }^{\infty }\frac{d\theta _{1}\ldots d\theta _{n}}{%
n!(2\pi )^{n}}\left\langle \mathcal{O}(x)\mathcal{\,}Z_{\mu _{1}}^{\dagger
}(\theta _{1})\ldots Z_{\mu _{n}}^{\dagger }(\theta _{n})\right\rangle
\left\langle Z_{\mu _{n}}^{\dagger }(\theta _{n})\ldots Z_{\mu
_{1}}^{\dagger }(\theta _{1})\right| \\
&=&\left\langle \mathcal{O}(x)\right. \mathcal{\,}\Bbb{I\,}.
\end{eqnarray*}
Cancelling the vacuum in the first and last line, and noting that we can
replace the product of operators, which is left over also by its normal
ordered version, we obtain the expression defined originally in \cite{Lask} 
\begin{equation}
\tilde{\!\mathcal{O}}(x)=\!\!\!\!\sum\Sb n=1\ldots \infty  \\ \mu _{1}\ldots
\mu _{n}  \endSb \int\limits_{-\infty }^{\infty }\frac{d\theta _{1}\ldots
d\theta _{n}}{n!(2\pi )^{n}}F_{n}^{\mathcal{O}|\mu _{1}\ldots \mu
_{n}}:Z_{\mu _{n}}^{\dagger }(\theta _{n})\ldots Z_{\mu _{1}}^{\dagger
}(\theta _{1}):e^{-i\sum\limits_{j}p_{j}\cdot x}.  \label{crap}
\end{equation}
Hence this field is simply an inversion of (\ref{1}). From its very
construction it is clear that $\tilde{\mathcal{O}}(x)$ is a meaningful field
in the weak sense, that is acting on an in-state we will recover by
construction the form factor related to $\mathcal{O}(x)$. In addition, one
may also construct the well-known expression of the two-point correlation
function expanded in terms of form factors, as stated in \cite{Lask}.
However, it is also clear that $\tilde{\mathcal{O}}(x)\neq \mathcal{O}(x),$
simply by comparing (\ref{crap}) and the explicit expressions for some local
fields occurring in the free fermionic theory, e.g. (\ref{aux}), (\ref{aux1}%
). The reason is that acting on an in-state with the latter expressions the
form factors are generated in a non-trivial Wick contraction procedure,
whereas when doing the same with (\ref{crap}) the Wick contractions will be
trivial. Therefore general statements and conclusions drawn from an analysis
based on the expression for $\tilde{\mathcal{O}}(x)$ should be taken with
care. It is also needless to say that from a practical point of view the
expression (\ref{crap}) is rather empty, since the expressions of the form
factors $F_{n}^{\mathcal{O}|\mu _{1}\ldots \mu _{n}}(\theta _{1}\ldots
\theta _{n})$ themselves are usually not known and their determination is in
general a quite non-trivial task. In \cite{Lask,Notas,Bert} the integration
in the formula (\ref{crap}) is a rather artificial contour integration which
takes care about analytic continuations of values of $i\pi $. This does not
seem to be a fundamental feature, since it remains completely obscure how to
incorporate bound states in this manner.

\section{The Federbush Model}

The Federbush model \cite{Feder} was proposed forty years ago as a prototype
for an exactly solvable quantum field theory which obeys the Wightman axioms 
\cite{Wight}. It contains two different massive particles $\Psi _{1}$ and $%
\Psi _{2}$. A special feature of this model is that the related vector
currents $J_{\alpha }^{\mu }=\bar{\Psi}_{\alpha }\gamma ^{\mu }\Psi _{\alpha
}$, $\alpha \in \{1,2\}$, whose analogues occur squared in the massive
Thirring model, enter the Lagrangian density of the Federbush model in a
parity breaking manner 
\begin{equation}
\mathcal{L}_{\text{F}}=\sum\nolimits_{\alpha =1,2}\bar{\Psi}_{\alpha
}(i\gamma ^{\mu }\partial _{\mu }-m_{\alpha })\Psi _{\alpha }-2\pi \lambda
\varepsilon _{\mu \nu }J_{1}^{\mu }J_{2}^{\nu }\,  \label{LFeder}
\end{equation}
due to the presence of the Levi-Civita pseudotensor $\varepsilon $. The
scattering matrix was found to be \cite{Wight,STW} 
\begin{equation}
S^{\text{FB}}=-\left( 
\begin{array}{cccc}
1 & 1 & e^{-2\pi i\lambda } & e^{2\pi i\lambda } \\ 
1 & 1 & e^{2\pi i\lambda } & e^{-2\pi i\lambda } \\ 
e^{2\pi i\lambda } & e^{-2\pi i\lambda } & 1 & 1 \\ 
e^{-2\pi i\lambda } & e^{2\pi i\lambda } & 1 & 1
\end{array}
\right) .  \label{SFeder}
\end{equation}
For the rows and columns we adopt here the ordering $\{1,\bar{1},2,\bar{2}\}$%
. In close relation to the free fermionic theory one may also introduce the
analogue fields to the disorder and order fields in the Federbush model 
\begin{eqnarray}
\Phi _{\alpha }^{\lambda }(x) &=&\mathbf{:}\exp [\Omega _{\alpha }^{\lambda
}(x)]\mathbf{:}=\vdots \exp [-2\sqrt{\pi }i\lambda \phi _{\alpha }(x)]\vdots
\label{ttt} \\
\Sigma _{\alpha }^{\lambda }(x) &=&\mathbf{:}\!\!\int \frac{dp_{\alpha }^{1}%
}{2\pi p_{\alpha }^{0}}(a_{\alpha }(p)e^{-ip_{\alpha }\cdot x}+a_{\bar{\alpha%
}}^{\dagger }(p)e^{ip_{\alpha }\cdot x})\,\Phi _{\alpha }^{\lambda }(x)%
\mathbf{:},  \label{problemilla}
\end{eqnarray}
where the $\kappa $-function related to $\Omega $ is 
\begin{equation}
\hat{\kappa}^{1}(\theta ,\theta ^{\prime })=-\hat{\kappa}^{2}(-\theta
,-\theta ^{\prime })=\frac{i\sin (\pi \lambda )e^{-\lambda (\theta -\theta
^{\prime })}}{2\cosh \frac{1}{2}(\theta -\theta ^{\prime })}.  \label{om}
\end{equation}
The last equality in (\ref{ttt}) was found by Lehmann and Stehr \cite{LS},
who showed the remarkable fact that the operator $\Phi _{\alpha }^{\lambda
}(x)$ can be viewed in two equivalent ways. On one hand it can be defined
through triple ordered free Bosons $\phi _{\alpha }(x)$, defined as $\mathbf{%
\vdots }e^{\kappa \phi }\mathbf{\vdots }=e^{\kappa \phi }/\left\langle
e^{\kappa \phi }\right\rangle $ for $\kappa $ being some constant, and on
the other hand by means of a conventional fermionic Wick ordered expression.
We compute \cite{CF5} the following equal time exchange relations for $%
\alpha ,\beta =1,2$%
\begin{eqnarray}
\psi _{\alpha }(x)\Phi _{\beta }^{\lambda }(y) &=&\Phi _{\beta }^{\lambda
}(y)\psi _{\alpha }(x)\,e^{2\pi i(-1)^{\beta }\lambda \delta _{\alpha \beta
}\Theta (x^{1}-y^{1})}\,,  \label{antic} \\
-\psi _{\alpha }(x)\Sigma _{\beta }^{\lambda }(y) &=&\Sigma _{\beta
}^{\lambda }(y)\psi _{\alpha }(x)\,e^{2\pi i(-1)^{\beta }\lambda \delta
_{\alpha \beta }\Theta (x^{1}-y^{1})}\,,  \label{345} \\
\Phi _{\alpha }^{\lambda }(x)\Phi _{\beta }^{\lambda }(y) &=&\Phi _{\beta
}^{\lambda }(y)\Phi _{\alpha }^{\lambda }(x) \\
\Sigma _{\alpha }^{\lambda }(x)\Sigma _{\beta }^{\lambda }(y) &=&\Sigma
_{\beta }^{\lambda }(y)\Sigma _{\alpha }^{\lambda }(x)\,e^{2\pi i(-1)^{\beta
}\lambda \delta _{\alpha \beta }}\,\,.  \label{fucked}
\end{eqnarray}
where $\Theta (x)$ is the Heavyside step function. With the relevant
exchange relations at our disposal, we can, according to (8), read off the
factors of local commutativity for the operators under consideration 
\begin{equation}
\gamma _{\alpha }^{\Phi _{\beta }^{\lambda }}=-\gamma _{\alpha }^{\Sigma
_{\beta }^{\lambda }}=e^{2\pi i(-1)^{\beta }\lambda \delta _{\alpha \beta
}}\quad \text{and\quad }\gamma _{\bar{\alpha}}^{\Phi _{\beta }^{\lambda
}}=-\gamma _{\bar{\alpha}}^{\Sigma _{\beta }^{\lambda }}=e^{-2\pi
i(-1)^{\beta }\lambda \delta _{\alpha \beta }}\quad .\,  \label{carallo}
\end{equation}

Proceeding again in the same way as in the previous section, we obtain as
closed expressions for the n-particle form factors 
\begin{eqnarray}
F_{2n}^{\Phi _{1}^{\lambda }|n\times \bar{1}1}(\bar{x}_{1},x_{2}\ldots \bar{x%
}_{2n-1},x_{2n})=(-1)^{n}F_{2n}^{\Phi _{2}^{-\lambda }|n\times \bar{2}2}(%
\bar{x}_{1},x_{2}\ldots \bar{x}_{2n-1},x_{2n})&=&  \nonumber \\
F_{2n}^{\Phi _{\bar{1}}^{-\lambda }|n\times \bar{1}1}(\bar{x}%
_{1},x_{2}\ldots \bar{x}_{2n-1},x_{2n})=(-1)^{n}F_{2n}^{\Phi _{\bar{2}%
}^{\lambda }|n\times \bar{2}2}(\bar{x}_{1},x_{2}\ldots \bar{x}%
_{2n-1},x_{2n})&=&  \nonumber \\
i^{n}2^{n-1}\sin ^{n}(\pi \lambda )\sigma _{n}(\bar{x}_{1}\ldots \bar{x}%
_{2n-1})^{\lambda +\frac{1}{2}}\sigma _{n}(x_{2}\ldots x_{2n})^{\frac{1}{2}%
-\lambda }\mathcal{B}_{n,n} ,&&  \label{chocho}
\end{eqnarray}
\begin{eqnarray}
\tilde{F}_{2n+1}^{\Sigma _{1}^{\lambda }|1(n\times \bar{1}1)}(\theta
_{1},\ldots ,\theta _{2n+1})&=&(-1)^{n}\tilde{F}_{2n+1}^{\Sigma
_{2}^{-\lambda }|2(n\times \bar{2}2)}(\theta _{1},\ldots ,\theta _{2n+1})= 
\nonumber \\
\tilde{F}_{2n+1}^{\Sigma _{\bar{1}}^{-\lambda }|1(n\times \bar{1}1)}(\theta
_{1},\ldots ,\theta _{2n+1})&=&(-1)^{n}\tilde{F}_{2n+1}^{\Sigma _{\bar{2}%
}^{\lambda }|2(n\times \bar{2}2)}(\theta _{1},\ldots ,\theta _{2n+1})= \frac{%
\sin ^{n}(\pi \lambda ) }{2}  \nonumber \\
\frac{ (2i)^{n} \sigma _{n}(\bar{x}_{2}\ldots \bar{x}_{2n})^{\lambda +\frac{1%
}{2}} }{ \sigma _{n}(x_{1}\ldots x_{2n+1})^{\lambda -\frac{1}{2}}}
\prod\limits_{1\leq i<j\leq n} \!\!\!\!\! &&\!\!\!\!\! (\bar{x}_{2i}-\bar{x}%
_{2j})\sum_{k}\frac{i^{k+1}\prod\limits_{j<l;j,l\neq k}(x_{j}-x_{l})}{%
(x_{k})^{\frac{1}{2}-\lambda }\,\prod\limits_{j\neq k}\prod\limits_{l}(x_{j}+%
\bar{x}_{l})}\,.  \label{85}
\end{eqnarray}
We may now convince ourselves, that the expressions for $F_{2n}^{\Phi
_{\alpha }^{\lambda }|n\times \bar{\alpha}\alpha }$ indeed satisfy the
consistency equations (\ref{W1})-(\ref{kin}). However, the expressions of $%
\tilde{F}_{2n+1}^{\Sigma _{\alpha }^{\lambda }|\alpha (n\times \bar{\alpha}%
\alpha )}$ only satisfy the consistency equations (\ref{W1})-(\ref{kin}) for 
$\lambda =1/2$. This reflects the very important fact that $\Sigma _{\alpha
}^{\lambda }(x)$ is only a mutually local operator for this value of $%
\lambda $, see equation (\ref{fucked}), unlike $\Phi_\alpha^\lambda(x)$
which is mutually local for all value of $\lambda$. Thus, the equations (\ref
{W1})-(\ref{kin}) select out solutions corresponding to operators which are
mutually local.

The form factors related to the trace of the energy-momentum tensor turn out
to be the same as the ones for the complex free Fermion.

\section{Momentum space cluster properties}

Cluster properties in space, i.e. the observation that far separated
operators do not interact, are quite familiar in quantum field theories \cite
{Wich} for a long time. In 1+1 dimensions a similar property has also been
noted in momentum space. It states that whenever some of the rapidities, say 
$\kappa $, are shifted to plus or minus infinity, the $n$-particle form
factor related to a local operator $\mathcal{O}$ factorizes into a $\kappa $
and an ($n-\kappa $)-particle form factor which are possibly related to
different types of operators $\mathcal{O}^{\prime }$ and $\mathcal{O}%
^{\prime \prime }$. This type of behaviour has been analyzed explicitly for
several specific models \cite{Zamocorr,KM,Smir2,CF1}. The possibility of
non-self-clustering, i.e. $\mathcal{O}\neq \mathcal{O}^{\prime }\neq 
\mathcal{O}^{\prime \prime }$, was conjectured for the first time in \cite
{KM} and the first explicit examples which confirm this were found in \cite
{CF1}. For self-clustering and a purely bosonic case this behaviour can be
explained perturbatively by means of Weinberg's power counting theorem, see
e.g. \cite{BK2} for an explicit reasoning on this issue. Non-self-clustering
still lacks an explanation at present. The cluster property serves not only
a consistency check for possible solutions of (\ref{W1})-(\ref{kin}), but
also as a construction principle for new solutions, e.g. \cite{CF1}.

An interesting operator related property which the form factors satisfy is
the momentum space cluster decomposition 
\begin{equation}
\lim_{\Delta \rightarrow \infty }F_{k+l}^{\mathcal{O}}(\theta _{1}\dots
\theta _{k},\theta _{k+1}+\Delta \dots \theta _{k+l}+\Delta )=F_{k}^{%
\mathcal{O^{\prime }}}(\theta _{1}\dots \theta _{k})F_{l}^{\mathcal{%
O^{\prime \prime }}}(\theta _{k+1}\dots \theta _{k+l})\;,
\label{eq: cluster}
\end{equation}
Writing instead of the matrix elements only the operators, we obtained \cite
{CF5} formally the following decomposition 
\begin{equation}
\Phi _{\alpha }^{\lambda }\longrightarrow \Phi _{\alpha }^{\lambda }\times
\Phi _{\alpha }^{\lambda }\quad \quad \sigma _{\alpha }\longrightarrow
\QATOPD\{ . {\mu _{\alpha }\times \sigma _{\alpha }}{\mu _{\bar{\alpha}%
}\times \sigma _{\alpha }}\quad \quad \mu _{\alpha }\longrightarrow
\QATOPD\{ . {\mu _{\alpha }\times \mu _{\alpha }}{\sigma _{\alpha }\times
\sigma _{\bar{\alpha}}}  \label{nonself}
\end{equation}
together with the equations for $\alpha \leftrightarrows \bar{\alpha}$. This
means the stated operator content closes consistently under the action of
the cluster decomposition operators. We also observe that
non-self-clustering, i.e. $\mathcal{O}\neq \mathcal{O}^{\prime }\neq 
\mathcal{O}^{\prime \prime }$, is possible. Unlike the self-clustering,
which can be explained for the bosonic case with the help of Weinberg's
power counting argument, this property is not yet understood from general
principles.

\section{Lie algebraically coupled Federbush models}

The Federbush model as investigated in the previous section only contains
two types of particles. In this section we propose a new Lagrangian, which
admits a much larger particle content. The theories are not yet as complex
as the homogeneous sine-Gordon (HSG) models, but they can also be obtained
from them in a certain limit such that they will always constitute a
benchmark for these class of theories.

Let us consider $\ell \times \tilde{\ell}$-real (Majorana) free Fermions $%
\psi _{a,j}(x)$, now labeled by two quantum \ numbers $1\leq a\leq \ell $, $%
1\leq j\leq \tilde{\ell}$ and described by the Dirac Lagrangian density $%
\mathcal{L}_{\text{FF}}$. We perturb this system with a bilinear term in the
vector currents $J_{a,j}^{\mu }=\bar{\Psi}_{a,j}\gamma ^{\mu }\Psi _{a,j}$ 
\begin{equation}
\mathcal{L}_{\text{CF}}=\sum_{a=1}^{\ell }\sum_{j=1}^{\tilde{\ell}}\bar{\Psi}%
_{a,j}(i\gamma ^{\mu }\partial _{\mu }-m_{a,j})\Psi _{a,j}-\frac{1}{2}\pi
\varepsilon _{\mu \nu }\sum_{a,b=1}^{\ell }\sum_{j,k=1}^{\tilde{\ell}%
}J_{a,j}^{\mu }J_{b,k}^{\nu }\Lambda _{ab}^{jk}\,\,,  \label{CAF}
\end{equation}
and denote the new fields in $\mathcal{L}_{\text{CF}}$ by $\Psi _{a,j}$.
Furthermore, we introduced $\ell ^{2}\times \tilde{\ell}^{2}$ dimensional
coupling constant dependent matrix $\Lambda _{ab}^{jk}$, whose further
properties we leave unspecified at this stage. We computed \cite{CF5} the
related S-matrix to 
\begin{equation}
S_{ab}^{jk}=-e^{i\pi \Lambda _{ab}^{jk}}\,\,.  \label{S}
\end{equation}
where due to the crossing and unitarity relations we have the constraints 
\begin{equation}
\Lambda _{ab}^{jk}=-\Lambda _{ba}^{kj}+2\Bbb{Z}\quad \quad \text{and\qquad }%
\Lambda _{ab}^{jk}=\Lambda _{\bar{b}a}^{\bar{k}j}+2\Bbb{Z\,}\,  \label{con}
\end{equation}
on the constants $\Lambda $. Taking $\Lambda _{ab}^{jk}=2\lambda
_{ab}\varepsilon _{jk}\tilde{I}_{jk}K_{a\bar{b}}^{-1}$, with $K,I$ being the
Cartan and incidence matrix, respectively, provides the limit of the
HSG-models.

\section{The ultraviolet limit}

The ultraviolet Virasoro central charge of the theory itself can be computed
from the knowledge of the form factors of the trace of the energy-momentum
tensor \cite{ZamC} by means of the expansion 
\begin{equation}
c_{\text{uv}}=\sum_{n=1}^{\infty }\sum_{\mu _{1}\ldots \mu _{n}}\frac{9}{%
n!(2\pi )^{n}}\int\limits_{-\infty }^{\infty }\ldots \int\limits_{-\infty
}^{\infty }\frac{d\theta _{1}\ldots d\theta _{n}\left| F_{n}^{T_{\;\;\mu
}^{\mu }|\mu _{1}\ldots \mu _{n}}(\theta _{1},\ldots ,\theta _{n})\right|
^{2}}{\left( \sum_{i=1}^{n}m_{\mu _{i}}\cosh \theta _{i}\right) ^{4}}\,.
\label{ccc}
\end{equation}
In a similar way one may compute the scaling dimension of the operator $%
\mathcal{O}$ from the knowledge of its n-particle form factors \cite{DSC} 
\begin{eqnarray}
\Delta _{\text{uv}}^{\!\!\mathcal{O}} &=&-\frac{1}{2\left\langle \mathcal{O}%
\right\rangle }\sum_{n=1}^{\infty }\sum_{\mu _{1}\ldots \mu
_{n}}\int\limits_{-\infty }^{\infty }\ldots \int\limits_{-\infty }^{\infty }%
\frac{d\theta _{1}\ldots d\theta _{n}}{n!(2\pi )^{n}\left(
\sum_{i=1}^{n}m_{\mu _{i}}\cosh \theta _{i}\right) ^{2}}  \nonumber \\
&&\times F_{n}^{T_{\;\;\mu }^{\mu }|\mu _{1}\ldots \mu _{n}}(\theta
_{1},\ldots ,\theta _{n})\,\left( F_{n}^{\mathcal{O}|\mu _{1}\ldots \mu
_{n}}(\theta _{1},\ldots ,\theta _{n})\,\right) ^{\ast }\,\,\,.
\label{dcorr}
\end{eqnarray}
In general the expressions (\ref{ccc}) and (\ref{dcorr}) yield the
difference between the corresponding infrared and ultraviolet values, but we
assumed here already that the theory is purely massive such that the
infrared contribution vanishes. Evaluating these formulae, we obtain 
\begin{equation}
c_{\text{uv}}=2\qquad \text{and\qquad }\Delta _{\text{uv}}^{\!\!\mu _{\alpha
}}=\Delta _{\text{uv}}^{\!\!\mu _{\bar{\alpha}}}=\frac{1}{16}\,.  \label{16}
\end{equation}
for the complex free Fermion and 
\begin{equation}
c_{\text{uv}}=2\qquad \text{and\qquad }\Delta _{\text{uv}}^{\!\!\Phi
_{\alpha }^{\lambda }}=\Delta _{\text{uv}}^{\!\!\Phi _{\alpha }^{\lambda }}=%
\frac{\lambda ^{2}}{4}\,.  \label{lam}
\end{equation}
for the Federbush model Note, that $\Delta _{\text{uv}}^{\!\!\Phi _{\alpha
}^{1/2}}=\Delta _{\text{uv}}^{\!\!\Phi _{\alpha }^{1/2}}=1/16$, which is the
limit to the complex free Fermion. Yet more support for the relation between
the $SU(3)_{3}$-HSG model and the Federbush model comes from the analysis of 
$\lambda =2/3$, for which the $SU(3)_{3}$-HSG S-matrix is related to the one
of the Federbush model. In that case we obtain from (\ref{lam}) the values $%
\Delta _{\text{uv}}^{\!\!\Phi _{\alpha }^{2/3}}=\Delta _{\text{uv}%
}^{\!\!\Phi _{\alpha }^{2/3}}=1/9$, which is a conformal dimension occurring
in the $SU(3)_{3}$-HSG model. Thus precisely at the value of the coupling
constant of the Federbush model at which the $SU(3)_{3}$-HSG S-matrix
reduces to the $S^{\text{FB}}$, the operator content of the two models
overlaps.

\section{Conclusions}

We summarize our main results:

We computed explicitly closed formulae for the n-particle form factors of
the complex free Fermion and the Federbush model related to various
operators.

We carried out this computations in two alternative ways: On the one hand,
we represent explicitly the field content (\ref{aux}) as well as the
particle creation operators (\ref{real}) in terms of fermionic Fock
operators (\ref{ferm}) and computed thereafter directly the corresponding
matrix elements. On the other hand we verified that these expressions
satisfy the form factor consistency equations only when the operators under
consideration are mutually local, i.e. satisfying (\ref{loc}). It is crucial
that the consistency equations contain the factor of local commutativity $%
\gamma _{\mu }^{\mathcal{O}}$ as defined in (\ref{local}). Our analysis
strongly suggest that \textit{the form factor consistency equations select
out operators, which are mutually local in the sense of (\ref{loc}).}

We carried out this computations in two alternative ways: On the one hand,
we represent explicitly the field content (\ref{aux}) as well as the
particle creation operators (\ref{real}) in terms of fermionic Fock
operators (\ref{ferm}) and computed thereafter directly the corresponding
matrix elements. On the other hand we verified that these expressions
satisfy the form factor consistency equations only when the operators under
consideration are mutually local, i.e. satisfying (\ref{loc}). This can
already be seen for the free Fermion, for which we could have also computed
the matrix element of the field $\Phi _{\alpha }^{\lambda }(x)$. In that
context one observes that only for $\lambda =1/2$ the resulting function $%
\tilde{F}$ solves the consistency equations (\ref{W1})-(\ref{kin}). We
observed a similar phenomenon in the Federbush model. Whereas the matrix
elements of the field $\Sigma _{\alpha }^{\lambda }(x)$ can be computed in a
closed form for generic values of $\lambda $, they become only meaningful
form factors for $\lambda =1/2$, that is when the field becomes local. This
means it is crucial that the consistency equations contain the factor of
local commutativity $\gamma _{\mu }^{\mathcal{O}}$ as defined in (\ref{local}%
), which we computed from first principles with the help of (\ref{antic})-(%
\ref{fucked}).

Our analysis strongly suggest that \textit{the form factor consistency
equations select out operators, which are mutually local in the sense of (%
\ref{loc}).} To establish this in complete generality still remains an open
issue. We have expressed our criticism on the analysis carried out in \cite
{Lask} in section 4. Further arguments which support this statement for
specific situations can be found in \cite{Smir,Thom}. These type of analysis
do not include the essential factor $\gamma $ and the latter one does not
allow higher order poles in the scattering matrix, which still excludes the
majority of know diagonal theories.

Our solutions turned out to decompose consistently under the momentum space
cluster property. This computations constitute next to the ones in \cite
{CFK,CF1} the first concrete examples of non-self-clustering, i.e. $\mathcal{%
O\rightarrow O}^{\prime }\times \mathcal{O}^{\prime \prime }$ in the sense
of (\ref{nonself}).

Further support for the identification of the solutions of (4)-(7) with a
specific operator was given by an analysis of the ultraviolet limit.

We demonstrated how the scattering matrix of the Federbush model can be
obtained as a limit of the $SU(3)_{3}$-HSG scattering matrix. This
``correspondence'' also holds for the central charge, which equals $2$ in
both cases, and the scaling dimension of the disorder operator at a certain
value of the coupling constant.

We proposed a Lie algebraic generalization of the Federbush models, by
suggesting a new type of Lagrangian. We evaluate from first principles the
related scattering matrices, which can also be obtained in a certain limit
from the HSG-models.

We expect that the construction of form factors by means of free fermionic
Fock fields can be extended to other models by characterizing further the
function $\kappa$.

\medskip
\noindent \textbf{Acknowledgments: } We are grateful to the Deutsche
Forschungsgemeinschaft (Sfb288),  INTAS project 99-01459 and the organisers
of the APCTP symposium for financial support. We would like to thank  the organisers 
for  their efforts and hospitality. We are specially indebted to Ed Corrigan, Tracey Dart, Chris Eilbeck, 
Mo-Lin Ge, Tigran Hakobyan, Li Jun, Tetsuji Miwa, Jacques Perk, 
Ara Sedrakyan and Robert Weston.

\end{document}